\documentclass{article}

\usepackage{arxiv}

\usepackage[utf8]{inputenc}
\usepackage[T1]{fontenc}
\usepackage{amsmath,amssymb,bm}
\usepackage{graphicx}
\usepackage{booktabs}
\usepackage{url}
\usepackage{cite}
\usepackage{microtype}
\usepackage[colorlinks,urlcolor=blue,linkcolor=blue,citecolor=blue]{hyperref}
\expandafter\def\expandafter\UrlBreaks\expandafter{\UrlBreaks\do\/\do\*\do\-\do\~\do\'\do\"\do\-}

\title{More Agents, More Reliable? System-Level Reliability of Multi-Agent LLMs under Operational-Profile Uncertainty}

\author{
  Burkay Tun\c{c}t\"urk\thanks{Corresponding author.} \\
  University of Hull, Hull, HU6 7RX, UK \\
  Bilkent University, Ankara, 06800, T\"urkiye \\
  \texttt{burkay.tuncturk@gmail.com} \\
  \And
  Koorosh Aslansefat \\
  University of Hull, Hull, HU6 7RX, UK \\
  \texttt{K.Aslansefat@hull.ac.uk} \\
  \AND
  Yiannis Papadopoulos \\
  University of Hull, Hull, HU6 7RX, UK \\
  \texttt{Y.I.Papadopoulos@hull.ac.uk} \\
  \And
  Bao Nguyen \\
  University of Leeds, Leeds, LS2 9JT, UK \\
  \texttt{b.nguyen@leeds.ac.uk} \\
}

\begin{document}
\maketitle
\begin{NoHyper}\renewcommand{\thefootnote}{}\footnotetext{This work has been submitted to the IEEE for possible publication. Copyright may be transferred without notice, after which this version may no longer be accessible.}\end{NoHyper}

\begin{abstract}
Teams increasingly build software from cooperating large language model (LLM) agents, often assuming more agents mean more reliability. Yet the usual evidence, average benchmark accuracy, ignores the input mix a deployed system meets and the rule merging the agents' answers. Building on HIP-LLM, a hierarchical imprecise-probability assessment of single LLMs, we propose a framework combining it with a Bayesian network of the system's failure logic for multi-agent systems. Agent failure-probability intervals propagate through this network, where the combining rule can itself fail, into a reliability envelope showing which evidence limits it. In a pipeline extracting regulated chemical-hazard statements, majority voting over three models protected against the weakest but did not clearly exceed the best, and a voter failing on 5\% of demands would largely erase its advantage over the middle model. More agents yield a more defensible reliability claim, not automatically a more reliable system.
\end{abstract}


\section{Introduction}

Software teams increasingly assemble large language models (LLMs) into multi-agent systems, in which several specialized agents divide a task and pass intermediate results to one another. Such systems are now entering scientific workflows, and in chemistry they already plan and run physical experiments~\cite{boiko2023autonomous}. In these settings an incorrect output is not simply a poor answer but a failure in the dependability sense~\cite{avizienis2004basic}, and a decision that must later be defended to users, auditors, or regulators.

Assessment has not kept pace with use: reported performance is almost always average benchmark accuracy, while the ways multi-agent systems actually fail receive far less attention~\cite{cemri2025why}. Two systems with the same score can behave very differently once operational inputs depart from those used in testing~\cite{bishop2017deriving}. Software reliability engineering has long treated this as a question of the operational profile, the probability distribution over the kinds of demand a system receives in the field~\cite{musa1993operational}. The profile is itself estimated from limited field data, and ignoring that uncertainty makes an estimate look more precise than the evidence supports. The problem is sharpest for systems assembled from independently developed components, where the assessor controls the composition but not the usage~\cite{pietrantuono2020reliability}.

A multi-agent LLM system adds a difficulty of its own. In the workflow of Figure~\ref{fig:overview}(a), a failure observed at the output need not originate at the last agent. An upstream agent may pass on a result that is wrong but well formed, a conditional edge may end a demand that should have continued, and a downstream agent may mask an upstream error, pass it on, or turn it into a different one. Propagation also decides whether redundancy helps. Replicating the final agent cannot mask an error that every replica receives from the same upstream stage. Without a model of propagation, the benefit of redundancy is assumed and not assessed.

Existing approaches each cover part of this picture. Recent work propagates calibrated confidence through a Bayesian network of workflow stages~\cite{custers2026runtime,donaldson2026bayesian}, but each node carries a single value not only a failure mode, the combination rule is assumed not to fail, and a benchmark stands in for the operational profile. Hierarchical imprecise-probability assessment weights an envelope across the operational profile, but for a single model~\cite{aghazadeh2026hipllm}. Classical fault tolerance votes replicated outputs but assumes independent replicas and an infallible voter~\cite{dubrova2013fault}, both questionable for agents that share prompts and data. Hierarchically Performed Hazard Origin and Propagation Studies (HiP-HOPS) synthesize system fault trees from the architecture~\cite{papadopoulos1999hiphops}, but need point probabilities from data sheets that an LLM agent does not have.

This article fits these pieces into one practical method for checking whether an architectural change, such as adding redundant agents, actually makes a multi-agent LLM system more reliable. The interval failure estimates that HIP-LLM~\cite{aghazadeh2026hipllm} gives for each agent become the basic-event probabilities of a Bayesian network that encodes the system's failure logic in the manner of HiP-HOPS. The rule that combines agent outputs is part of that logic and is allowed to fail. The result is a system-level reliability envelope whose width can be traced to agent evidence, the operational profile, and the combination rule. We demonstrate the approach on one design change to a real pipeline: making its hazard-extraction agent redundant with triple modular redundancy (TMR). The pipeline returns the hazard statements that the European Chemicals Agency (ECHA) lists for a substance.

\section{Methodology}

\begin{figure}[!tp]
\centering
\includegraphics[width=\textwidth,height=0.85\textheight,keepaspectratio]{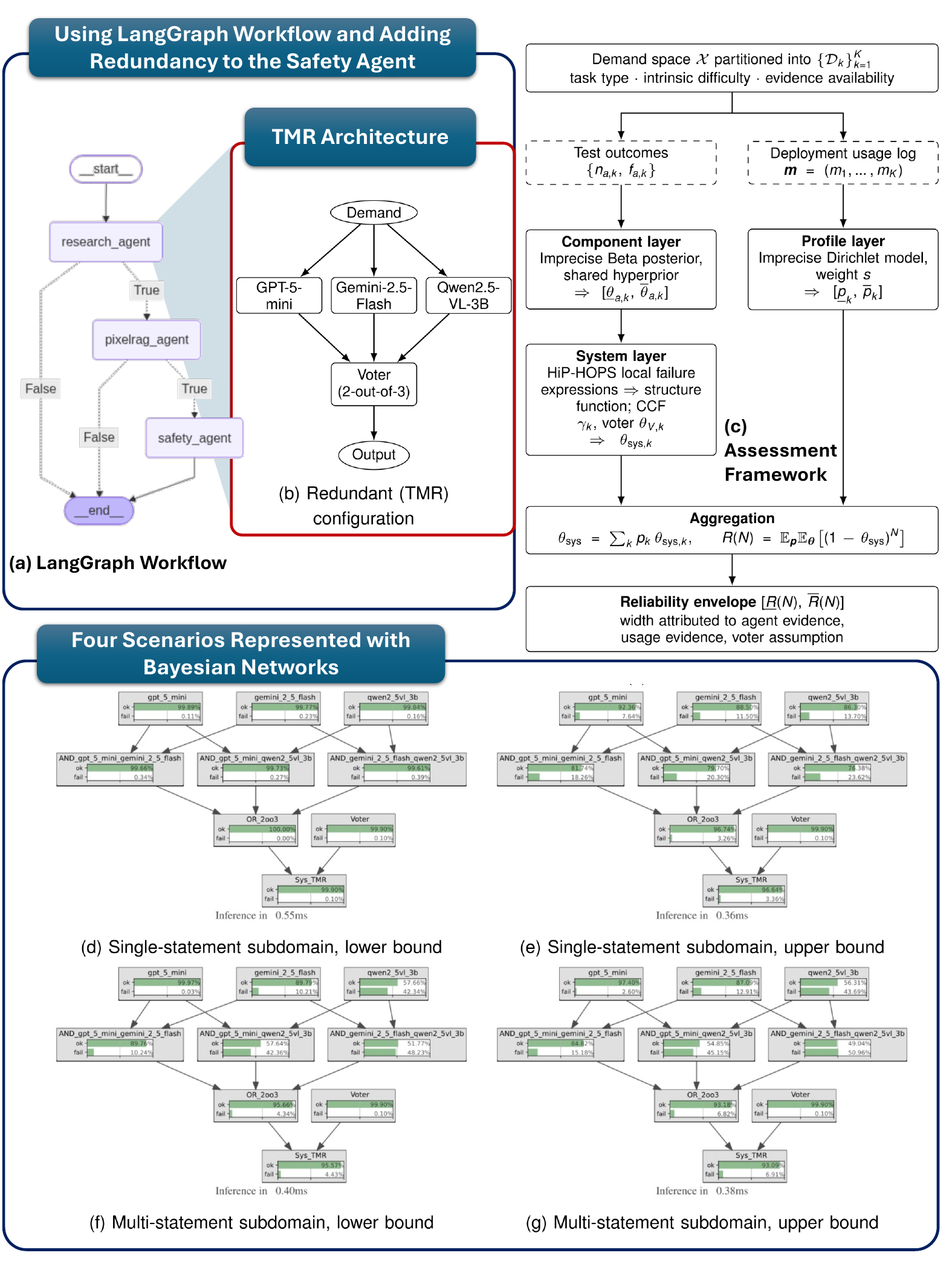}
\caption{Architectures and analysis. (a) Workflow diagram. (b) Evaluated TMR configuration. (c) Assessment framework. (d)--(g) Bayesian network of the TMR configuration at the lower and upper envelope ends for each subdomain. Layout and styling assisted by Claude Opus 5.5 (Anthropic).}
\label{fig:overview}
\end{figure}

The framework takes three inputs a team can collect, namely per-agent test outcomes, a usage log or declared operational profile, and the workflow's combination logic, and processes them in three layers (Figure~\ref{fig:overview}(c)): each agent's probability of failure on demand is assessed per subdomain under an imprecise prior, the combination rule is encoded as a structure function, and subdomain results are aggregated over an uncertain operational profile.

\subsection{Operational Profile: Definition and Estimation}

Following Aghazadeh-Chakherlou et al.~\cite{aghazadeh2026hipllm}, the assessor partitions the demand space into $K$ disjoint subdomains $\mathcal{D}_k$, chosen so that demands within a subdomain are exchangeable with respect to difficulty. A demand is one complete input to the workflow, in the case study one substance. The operational profile is the probability vector $\bm{p}=(p_1,\dots,p_K)$ with $p_k=\Pr(X\in\mathcal{D}_k)$. For an LLM agent the partition is built from features that predict difficulty, such as task type, input difficulty, and evidence availability.

Let a log of $M$ demands assign $m_k$ demands to $\mathcal{D}_k$. A single Dirichlet prior commits the assessor to a belief about usage that is rarely defensible when $M$ is small. Following the imprecise Dirichlet model~\cite{walley1996inferences}, we fix only the total prior weight $s$, which yields
\begin{equation}
\label{eq:idm_bounds}
\underline{\mathbb{E}}[p_k] = \frac{m_k}{s+M},
\qquad
\overline{\mathbb{E}}[p_k] = \frac{m_k + s}{s+M}.
\end{equation}
The interval narrows as usage evidence grows, and its width is carried into the final claim and not hidden by a convenient prior.

\subsection{Component Layer: Agent Reliability Within a Subdomain}

Let $\theta_{a,k}$ be the probability that agent $a$ fails on a demand drawn from $\mathcal{D}_k$, where failure is defined by the task oracle. Given $n_{a,k}$ test demands of which $f_{a,k}$ fail, the posterior is $\mathrm{Beta}(\alpha_k+f_{a,k},\,\beta_k+n_{a,k}-f_{a,k})$, with the hyperparameters drawn from a set instead of being fixed, so that the posterior mean failure probability ranges over an envelope $[\underline{\theta}_{a,k},\overline{\theta}_{a,k}]$. The envelope expresses prior imprecision; it is not a credible interval and does not include sampling uncertainty. In the reference implementation, the reliability $1-\theta_{a,k}$ in each subdomain is $\mathrm{Beta}(\mu\nu,\,(1-\mu)\nu)$, with $\mu$ and $\nu$ shared across subdomains, $\mu\sim\mathrm{Beta}(a,b)$ and $\nu\sim\mathrm{Gamma}(c,d)$ with rate $d$. The prior set is the default box $a,b\in[1,12]$, $c,d\in[1,25]$, sampled at 32 Latin-hypercube configurations of 1{,}200 posterior samples each; the envelope is the range of their posterior means. This shared hyperprior lets a subdomain with few observations borrow strength from the others without having its own evidence overwritten~\cite{aghazadeh2026hipllm}.

\subsection{System Layer: Composition as a Failure Expression}

Following HiP-HOPS~\cite{papadopoulos1999hiphops}, each agent is annotated with a local failure expression, and these are combined into a system-level expression. Here the expression for each configuration is written by hand and evaluated as a Bayesian network. In the redundant configuration, three replicas answer the same demand and a majority voter combines their answers H-code by H-code. The vote fails only if two replicas err on the same H-code: if one omits H1 and another omits H2, each code still gets two votes, but if both omit H1, the vote loses it. With three valid replicas, counting any two failed replicas as a system failure, as below, is therefore a conservative approximation, not the vote's exact failure logic. Writing $\Phi_k$ for the probability that at least two agents fail and $\theta_{V,k}$ for the probability that the voter itself fails,
\begin{equation}
\label{eq:tmr}
\theta_{\mathrm{sys},k} = \theta_{V,k} + (1-\theta_{V,k})\,\Phi_k .
\end{equation}
The first term is the failure floor imposed by the composition rule: no amount of agent redundancy reduces system failure below $\theta_{V,k}$. Architecturally, the combination rule is a single point of failure, so fail-safe design has to address it directly; adding agents cannot. Human oversight is one such option: escalating to a reviewer when the agents disagree takes those demands away from the voter, and the reviewer enters the structure function as a component with its own failure probability; this is not evaluated here.

The second term requires the joint failure distribution of the agents. Even agents built on different base models typically share a prompt template, an evidence source, and overlapping training data, so independence cannot simply be assumed~\cite{eckhardt1985theoretical}. We adopt a beta-factor decomposition~\cite{iec61508part6}. With $\theta^{\min}_k$ the failure probability of the strongest agent and $\gamma_k\in[0,1]$ the fraction of it attributable to a shared cause,
\begin{equation}
\label{eq:ccf_split}
\theta^{c}_{k} = \gamma_k\, \theta^{\min}_k,
\qquad
\theta^{i}_{a,k} = \frac{\theta_{a,k} - \theta^{c}_{k}}{1-\theta^{c}_{k}} .
\end{equation}
A shared failure defeats all three agents at once. Otherwise the residual parts act independently, and the network counts a failure when at least two do. At $\gamma_k = 1$ the redundant configuration can never be more reliable than its strongest agent, and setting $\gamma_k = \theta_{V,k} = 0$ recovers the textbook expression~\cite{dubrova2013fault}; classical treatments assume both away, whereas here they are sensitivity parameters.

\subsection{Aggregation and the Reliability Envelope}

The system probability of failure on demand and the reliability over $N$ future demands are
\begin{equation}
\label{eq:reliability}
\theta_{\mathrm{sys}} = \sum_{k=1}^{K} p_k \, \theta_{\mathrm{sys},k},
\qquad
R(N) = \mathbb{E}_{\bm{p}}\,\mathbb{E}_{\bm{\theta}}\big[(1-\theta_{\mathrm{sys}})^{N}\big].
\end{equation}
The form follows Aghazadeh-Chakherlou et al.~\cite{aghazadeh2026hipllm}, but $\theta_{\mathrm{sys}}$ now comes from the structure function, so the expectation carries the voter and common-cause terms. In practice, the network is evaluated at both ends of each component envelope, which bounds $\theta_{\mathrm{sys},k}$ because the two-out-of-three structure is monotone in each agent's failure probability, and the subdomain results are weighted by the operational profile. Table~\ref{tab:results} reports the plug-in value $(1-\theta_{\mathrm{sys}})^N$ at each end; because this function is convex in $\theta_{\mathrm{sys}}$, it slightly understates the expectation in \eqref{eq:reliability} and is therefore conservative.

\section{Experimental Evaluation}

The evaluation asks two questions: whether component-level test evidence for nondeterministic agents can be composed into a system-level reliability claim that agrees with the system's observed behavior, and whether that composition tells a designer something actionable about adopting redundancy.

\subsection{Case Study and Task Definition}

The case study is a LangGraph pipeline that takes a substance as a SMILES string and returns the hazard statements (H-codes) that ECHA lists for it. A research agent locates the substance's ECHA page, a capture agent renders the page into image tiles, and a vision-capable safety agent reads the tiles and returns the H-codes. We assess the safety agent, whose output is acted upon, and fix its input by running it on saved snapshots of the substances' ECHA CHEM pages (chem.echa.europa.eu), archived in the repository and captured on 17 August 2026. For six substances the listed classification is the harmonised classification under the CLP Regulation; for O-acetylsalicylic acid, paracetamol, and clozapine, which have none, it is the industry classification that ECHA displays (for clozapine, the majority notification). An output is correct only when its set of H-codes equals the ECHA record, not when the agents agree. The domain is used because it is regulated and safety-relevant, has an externally defined ground truth, and has inputs that stratify by difficulty; it is one instance of a pattern found well beyond chemistry, an LLM agent extracting structured facts from documents that a downstream process acts on. The nine substances, an ad hoc selection instead of a random or representative sample, are 1,1,1-trichloroethane, acetone, clozapine, colecalciferol, ethanol, methanol, nicotine, O-acetylsalicylic acid, and paracetamol. Their classifications are public and may appear in the models' training data; the prompt instructs the agent not to use general chemical knowledge, but this cannot be enforced, so part of the task may measure recall instead of reading of the page.

Two subdomains are used, substances with a single hazard statement ($\mathcal{D}_1$) and substances with two or more ($\mathcal{D}_2$), under an operational profile $\bm{p}=(0.20,\,0.80)$, declared instead of being estimated from a usage log, so \eqref{eq:idm_bounds} is not exercised here. Because this partition uses the ECHA record and not an input feature, assigning a deployment log to it, as \eqref{eq:idm_bounds} requires, would need an input-observable proxy such as molecular size; here it only stratifies test evidence, where the record is available. Nine substances were each submitted ten times to each of three safety-agent replicas built on GPT-5-mini, Gemini-2.5-Flash, and Qwen2.5-VL-3B (API identifiers gpt-5-mini and gemini-2.5-flash, with no dated snapshot pinned, and qwen2.5vl:3b served through Ollama; Gemini and Qwen at temperature 0, GPT-5-mini at its default settings), giving 270 calls, of which 3 were excluded as infrastructure errors (Gemini HTTP 503 errors that persisted over three attempts); unparseable responses were retried likewise. The fixed prompt, JSON output schema, and parser are in the repository. Repeated submission exposes the run-to-run nondeterminism of each agent, which a single benchmark pass would hide. The runs are not independent demands, however: with nine substances, two in $\mathcal{D}_1$ and seven in $\mathcal{D}_2$, the effective sample size lies between 9 and 267, yet the Beta update treats each call as independent, so the envelopes below are likely narrower than the evidence warrants. Qwen2.5-VL-3B was chosen as a small open-weight model of the kind teams host locally for cost or data protection, not to produce a spread of agent quality. Reliability is reported over $N=10$ future demands as an illustrative horizon. Other horizons follow from the $\theta_{\mathrm{sys}}$ column as $(1-\theta_{\mathrm{sys}})^N$; at the upper ends of their envelopes, reliability falls below one half after 11 demands for the redundant configuration and after 19 for the best agent.

\subsection{System Configurations}

Two configurations are assessed. In the simplex configuration, each agent alone sits on the output path, so the system fails exactly when it does. The redundant configuration, triple modular redundancy (TMR; Figure~\ref{fig:overview}(b)), presents the same page to all three replicas separately and keeps each H-code that a majority of the valid replicas report, so when a call is lost the two remaining replicas must agree. The comparison of interest is between the redundant system and the best individual agent, since a designer who declines redundancy must still choose one agent to deploy.

\begin{table}[!t]
\caption{Agent and system results. (a) Test outcomes (failures $f$ out of $n$ runs), posterior envelopes on the probability of failure on demand per subdomain, the envelope marginalized over the operational profile, and the reliability envelope over $N=10$ future demands. TMR rows use $\theta_V = 10^{-3}$; the propagated row is built only from the three agent rows, and the measured row fits the component layer to the observed voted outcomes as a check on the propagation. (b) Sensitivity of the propagated redundant configuration to the common-cause fraction $\gamma$ of \eqref{eq:ccf_split} and the voter failure probability $\theta_V$; entries are $[\underline{\theta}_{\mathrm{sys}},\overline{\theta}_{\mathrm{sys}}]$ marginalized over the declared profile.}
\label{tab:results}
\small
\setlength{\tabcolsep}{3pt}
\begin{tabular*}{\textwidth}{@{\extracolsep{\fill}}lcccccc@{}}
\multicolumn{7}{@{}l}{(a) Agent-level and system-level results}\\
\toprule
Configuration & $f/n$ in $\mathcal{D}_1$ & $f/n$ in $\mathcal{D}_2$ & Envelope in $\mathcal{D}_1$ & Envelope in $\mathcal{D}_2$ & $[\underline{\theta}_{\mathrm{sys}},\,\overline{\theta}_{\mathrm{sys}}]$ & $[\underline{R}(10),\,\overline{R}(10)]$ \\
\midrule
GPT-5-mini (best)         & 0/20 & 0/70  & $[.0011,\,.0764]$ & $[.0003,\,.0260]$ & $[.0005,\,.0361]$ & $[.6923,\,.9950]$ \\
Gemini-2.5-Flash (middle) & 0/18 & 7/69  & $[.0023,\,.1150]$ & $[.1021,\,.1291]$ & $[.0822,\,.1263]$ & $[.2592,\,.4241]$ \\
Qwen2.5-VL-3B (worst)     & 0/20 & 30/70 & $[.0016,\,.1370]$ & $[.4234,\,.4369]$ & $[.3390,\,.3769]$ & $[.0088,\,.0159]$ \\
\midrule
TMR, propagated & -- & -- & $[.0010,\,.0336]$ & $[.0443,\,.0691]$ & $[.0357,\,.0620]$ & $[.5273,\,.6952]$ \\
TMR, measured   & 0/20 & 1/70 & -- & -- & $[.0120,\,.0499]$ & $[.5994,\,.8863]$ \\
\bottomrule
\end{tabular*}

\vspace{8pt}
\begin{tabular*}{\textwidth}{@{\extracolsep{\fill}}lccccc@{}}
\multicolumn{6}{@{}l}{\makebox[0pt][l]{(b) Sensitivity of the redundant configuration (best agent alone: $[.0005,\,.0361]$; middle agent alone: $[.0822,\,.1263]$)}}\\
\toprule
$\gamma \backslash \theta_V$ & $0$ & $10^{-3}$ & $10^{-2}$ & $0.05$ & $0.10$ \\
\midrule
0.00 & $[.0347,\,.0611]$ & $[.0357,\,.0620]$ & $[.0443,\,.0705]$ & $[.0830,\,.1080]$ & $[.1312,\,.1550]$ \\
0.10 & $[.0347,\,.0618]$ & $[.0357,\,.0628]$ & $[.0444,\,.0712]$ & $[.0830,\,.1087]$ & $[.1312,\,.1556]$ \\
0.25 & $[.0347,\,.0631]$ & $[.0357,\,.0640]$ & $[.0444,\,.0724]$ & $[.0830,\,.1099]$ & $[.1313,\,.1568]$ \\
0.50 & $[.0348,\,.0654]$ & $[.0358,\,.0664]$ & $[.0445,\,.0748]$ & $[.0831,\,.1121]$ & $[.1313,\,.1589]$ \\
1.00 & $[.0349,\,.0714]$ & $[.0359,\,.0723]$ & $[.0446,\,.0807]$ & $[.0832,\,.1178]$ & $[.1314,\,.1642]$ \\
\bottomrule
\end{tabular*}
\end{table}

\subsection{Agent-Level Results}

In the multi-statement subdomain (Table~\ref{tab:results}(a)) the agents separate clearly: GPT-5-mini did not fail, Gemini-2.5-Flash failed on 7 of 69 runs, and Qwen2.5-VL-3B on 30 of 70. In the single-statement subdomain no agent failed, yet the upper bounds differ, in the same order as the agents' failures in $\mathcal{D}_2$. This follows from the shared hyperprior: an agent's bound in one subdomain depends on its failures in the other, so the $\mathcal{D}_1$ envelopes reflect the prior set and the $\mathcal{D}_2$ evidence and not failures observed in $\mathcal{D}_1$.

The component layer was also re-run at the substance level, where a substance fails if the agent fails on a majority of its runs (for every agent, equivalently on any run). No agent then fails in $\mathcal{D}_1$, and GPT-5-mini, Gemini-2.5-Flash, and Qwen2.5-VL-3B fail on 0, 1, and 3 of the seven substances in $\mathcal{D}_2$. The marginalized envelopes widen to $[.0038,\,.3351]$, $[.1187,\,.4177]$, and $[.3265,\,.5636]$, and the propagated TMR envelope to $[.0484,\,.4066]$. The voted outcomes fail on one substance under the any-failure rule (acetone, in 1 of 10 runs) and on none under the majority rule, giving measured envelopes of $[.1157,\,.4008]$ and $[.0039,\,.3222]$. The ordering is unchanged, but the envelopes now overlap, so at this sample size the evidence cannot separate the redundant configuration from the best agent.

\subsection{System-Level Results: With and Without Redundancy}

In the single-statement subdomain (Figure~\ref{fig:overview}(d)--(e)) no agent failed, so the gap between the two panels reflects the prior set and the $\mathcal{D}_2$ evidence, as noted above. In the multi-statement subdomain (panels f and g) the strongest agent rarely fails, so the network attributes system failure mainly to the two weaker agents failing on the same demand, which it counts as outvoting the strongest. This explains the result in Table~\ref{tab:results}(a): under this conservative approximation, majority voting protects against the weakest agent but does not clearly reach the reliability of the best agent alone. The voter adds its floor in every panel, but at $\theta_V = 10^{-3}$ this floor is small compared with the joint failure of the two weaker agents.

\subsection{Sensitivity Analysis}

Table~\ref{tab:results}(b) varies the voter failure probability $\theta_V$ and the common-cause factor $\gamma$. Raising $\gamma$ widens the envelope mainly at its upper end, while raising $\theta_V$ shifts the whole envelope upward. At $\theta_V = 0.05$ the lower bound of the redundant configuration reaches that of the middle agent, so beyond this point the voter, not the agents, drives the risk at the lower end of the envelope, and redundancy no longer improves even on the middle agent.

Moving from a fixed $p_1 = 0.20$ to $p_1 \in [0.05,\,0.35]$ widens the envelope of the redundant configuration from .0264 to .0382, while the part contributed by agent evidence stays at .0264. For this illustrative interval, the envelope is therefore limited more by agent evidence than by usage evidence.

\section{Discussion}

The central claim is compositional: reliability estimates for single agents can be carried, through a structure function that models the combination rule exactly or conservatively, into a defensible claim about the configuration that contains them, so component-level testing of nondeterministic agents can inform system-level validation.

\subsection{Propagated and Measured Envelopes Overlap}

Table~\ref{tab:results}(a) checks this claim: the propagated envelope uses only the replicas' separate evidence, under independence ($\gamma = 0$), and the measured one is fitted to the voted outcomes of the same runs, so this is a consistency check, not a prediction on held-out demands. The two overlap, but the measured envelope lies below the propagated one, so the propagation errs on the pessimistic side. The conservative approximation of the vote contributes little here: every Gemini-2.5-Flash failure returned no H-codes and every Qwen2.5-VL-3B failure omitted one, so a run in which both failed would also have failed the vote, although Gemini's failures come from a single substance. Instead, in $\mathcal{D}_2$ independence predicts about three such runs, but none occurred. This is weak evidence about correlation, because failures clustered by substance: with one and three failing substances out of seven, no overlap has probability 4/7 by chance. The vote masked 36 of the 37 single-replica failures. The only voted failure came when a Gemini call lost to an HTTP 503 error left two replicas and one omitted an H-code; on this fallback, which the network does not model, one omission is enough, so the approximation is not conservative there.

Keeping agent evidence, voter reliability, and correlation as separate terms in \eqref{eq:tmr} is what makes the claim transparent and accountable: an assessor can state not only how reliable the system is, but why, and which evidence would change that answer.

\subsection{Implications for Practice}

Three lessons from the case study translate into repeatable practices for teams that build multi-agent LLM software. First, compare a redundant configuration with the best single agent, not the average one. Second, test and monitor the combination rule like any other component, and model it at the granularity it works at, including its fallback when an agent is unavailable; here that fallback caused the only voted failure. Third, report reliability as an envelope tied to a stated operational profile, together with what limits its width; here, for an illustrative profile interval, it was agent evidence, which tells the team where further testing pays off. Once the failure logic is encoded, the assessment can be rerun whenever a model version, prompt, or usage mix changes and used as a release gate, much like a regression suite; automating that encoding would make it routine in operational governance.

\subsection{Scope of the Case Study}

Only one domain, one redundancy structure, and nine substances are used to exercise the composition, so the specific attributions above should be read as a demonstration of what the decomposition can surface, not as settled claims about these models. A larger study, with more substances and held-out demands, is needed to test whether the dominance of agent evidence, and the conservatism identified above, still hold.

\section{Future Work}

The attributions above rest on parameters the assessor supplies, $\gamma$ and $\theta_V$; the most direct extensions estimate them instead. The common-cause factor $\gamma$ could be estimated from the agents' disagreement pattern on held-out demands, so that replicas that fail together are flagged and monitored, and a failure log for the combination logic would turn $\theta_V$ into an estimate.

A second extension is automation: the HiP-HOPS-LLM tool can synthesize the failure logic from a LangGraph workflow, replacing the hand-built network, encoding the vote per H-code with its fallback instead of approximating it, and letting chains, conditional routing, and larger votes, which shift the failure floor and the influence of $\gamma$, be assessed. Domains without an externally fixed oracle would test how far these results generalize.

\section{Conclusion}
This article showed how a team can bound the reliability of a multi-agent LLM system from evidence it can collect, by combining hierarchical imprecise-probability assessment of each agent with a Bayesian network of the system's failure logic.

With one hazard-extraction agent made redundant by TMR, an envelope built from the three replicas' separate evidence agreed in order of magnitude with one fitted to the voted outcomes, and the decomposition showed that voting guarded against the weakest replica but did not clearly exceed the best one. Under its assumptions, the network approximates the implemented per-H-code vote conservatively instead of reproducing its exact failure logic; for the error modes observed the two coincide, but the network omits the two-replica fallback that caused the only voted failure, so how the combination rule is modeled is part of the claim. The answer to the question in the title is therefore conditional instead of being categorical: more agents compose into a more defensible reliability claim, not automatically into a more reliable system. For teams building AI-enabled software, the practical message is that trust in a multi-agent architecture should be earned from evidence about its agents, its combination rule, and its usage, each of which can be collected, monitored, and updated as the system evolves.

\section{Code Availability}
The source code, experiment notebooks, ECHA page snapshots, ground-truth data and raw results are publicly available at \url{https://github.com/Bruhkay/hip-llm-multiagent-reliability}. The HiP-HOPS-LLM tool used for structural synthesis is available at \url{https://github.com/koo-ec/HIP_HOPS_LLM}.

\bibliographystyle{IEEEtran}
\bibliography{references}

\end{document}